\documentclass[12pt]{article}
\usepackage[dvips]{graphicx}
\usepackage{amsmath,amssymb,amsfonts}

\begin{document}

\title{Kaon regeneration}
\author{V.I. Nazaruk\\
Institute for Nuclear Research of RAS, 60th October\\
Anniversary Prospect 7a, 117312 Moscow, Russia.*}

\date{}
\maketitle
\bigskip

\begin{abstract}
We have performed the calculation of kaon regeneration based on the exact solution of equations of motion. The result differs radically from the previous one.

\end{abstract}

\vspace{5mm}

\vspace{5mm}

\vspace{1cm}

*E-mail: nazaruk@inr.ru

\newpage
\setcounter{equation}{0}
\section{Introduction}
In the pioneer paper [1] a formalism of $K^0_{S}$-component regeneration has been considered. This approach was futher developed in [2,3]. The results of [1-3] were included in [4-7]. 

However, in [1-3] a system of non-coupled equations of motion has been considered instead of the coupled ones (see Eq. (7) below). In other words, the problem different from regeneration has been solved. In this connection an alternative calculation based on the perturbation theory has been proposed [8]. The similar approach is well studied for $n\bar{n}$ transitions in the medium [9-11]. Since the result obtained in Ref. [8] fundamentally differs from the previous one, in this paper we present the calculation based on the exact solution of equations of motion.
 
In Sect. 2 the exact wave functions are obtained. The probability of finding $K^0_{S}$ is calculated in Sect. 3. In Sect. 4 the comparison with the old calculations based on uncoupled equations is given. Section 5 contains the conclusion.

\section{Wave functions}
Let $K^0_{L}$ fall onto the plate at $t=0$. We use notations of Ref. [5]. Since 
\begin{equation}
\left.\mid\!K^0_L\right>=\left.(\mid\!K^0\right>+\left.\mid\!\bar{K}^0\right>)/\sqrt{2},
\end{equation}
the evolution of  $K^0_{L}$ in the medium is described by the following equation:
\begin{equation}
\left.\mid\!K'_L(t)\right>=(\left.\mid\!K^0(t)\right>+\left.\mid\!\bar{K}^0(t)\right>)/\sqrt{2}=[\left.\mid\!K^0\right>K^0(t)+\left.\mid\!\bar{K}^0\right>\bar{K}^0(t)]/\sqrt{2}.
\end{equation}
Here $\left.\mid\!K^0\right>$ and $\left.\mid\!\bar{K}^0\right>$ are the states of $K^0$ and $\bar{K}^0$, respectively; $K^0(t)$ and $\bar{K}^0(t)$ are the amplitudes of states (spatial wave functions) of  $K^0$ and $\bar{K}^0$, respectively. $K^0(t)$ and $\bar{K}^0(t)$ are calculated in $K^0,\bar{K}^0$ representation.

With these $K^0(t)$ and $\bar{K}^0(t)$ we revert to $K^0_{L},K^0_{S}$ representation:
\begin{eqnarray}
\left.\mid\!K'_L(t)\right>=\frac{1}{\sqrt{2}}[(\left.\mid\!K_L\right>+\left.\mid\!K_S\right>)K^0(t))/\sqrt{2}+
(\left.\mid\!K_L\right>-\left.\mid\!K_S\right>)\bar{K}^0(t))/\sqrt{2}]=\nonumber\\
\left.K_L(t)\mid\!K_L\right>+\left.K_S(t)\mid\!K_S\right>,  
\end{eqnarray}
where
\begin{eqnarray}
K_L(t)= \frac{1}{2}[K^0(t)+\bar{K}^0(t)],\nonumber\\
K_S(t)= \frac{1}{2}[K^0(t)-\bar{K}^0(t)].
\end{eqnarray}
$\mid \!K_S(t)\!\mid ^2$ is the probability of finding $K^0_{S}$. The value of $\mid \!K_S(t)\!\mid ^2$ is of
particular interest. 

Let us calculate $K^0(t)$ and $\bar{K}^0(t)$. The coupled equations for zero momentum $K^0$ and $\bar{K}^0$ in the medium are the following:
\begin{eqnarray}
(i\partial_t-M)K^0=\epsilon \bar{K}^0,\nonumber\\
(i\partial_t-(M+V))\bar{K}^0=\epsilon K^0,
\end{eqnarray}
where
\begin{eqnarray}
M=m_{K^0}+U_{K^0}-i\Gamma _{K^0}^d/2,\nonumber\\
V=(m_{\bar{K}^0}-m_{K^0})+(U_{\bar{K}^0}-U_{K^0})-(i\Gamma _{\bar{K}^0}^d/2-i\Gamma _{K^0}^d/2).
\end{eqnarray}
Here $\epsilon =(m_L-m_S)/2=\Delta m/2$ is a small parameter, $U_{K^0}$ and $U_{\bar{K}^0}$
are the potentials of $K^0$ and ${\bar{K}^0}$, $\Gamma _{K^0}^d$ and 
$\Gamma _{\bar{K}^0}^d$ are the decay widths of $K^0$ and ${\bar{K}^0}$, respectively.
 
The fundamental difference between our and previous calculations lies in the 
process model. For the previous old calculations [2] the starting equations are (see Eqs. (3) from [2]):
\begin{eqnarray}
(\partial_x-ink)K^0=0,\nonumber\\
(\partial_x-in'k)\bar{K}^0=0,
\end{eqnarray}
where $n$ and $n'$ are the indexes of refraction for $K^0$ and $\bar{K}^0$, 
respectively. In notations of Ref. [2] $K^0=\alpha $ and $\bar{K}^0=\alpha '$,
$K^0_{S}=\alpha _1$ and $K^0_{L}=\alpha _2$. In above-given Eq. (7) we substitute $K^0=(\alpha _1+ i\alpha _2)/ \sqrt{2}$, $\bar{K}^0=(\alpha _1-i\alpha _2)/ \sqrt{2}$ and include the effect of the weak interactions as in [2]. We obtain Eq. (5) and result (6) from Ref. [2].

So the starting equations (3) from [2] are non-coupled. There is no off-diagonal mass $\epsilon =(m_L-m_S)/2$. This is a fundamental defect.
The non-coupled equations exist only for the stationary states and don't exist for $K^0$ and $\bar{K}^0$. Equations (5) given above 
should be considered, not (7). For $n\bar{n}$ transitions in the medium [9-13] the coupled 
equations type of (5) for the non-stationary states are solved as well as for any $ab$-oscillations.

The exact solutions of (5) have the form
\begin{eqnarray}
K^0(t)=C_1e^{S_1t}+C_2 e^{S_2t},\nonumber\\
\bar{K}^0(t)= \frac{1}{2\epsilon }[C_1(V-p) e^{S_1t}+C_2(V+p) e^{S_2t}],\nonumber\\
S_{1,2}=\frac{i}{2}(-(2M+V)\pm p),\nonumber\\
p=\sqrt{4\epsilon ^2+V^2}.
\end{eqnarray}
Substituting (8) into (4) we obtain
\begin{eqnarray}
K_L(t)=C_3e^{S_1t}(1+\frac{V-p}{2\epsilon }) +C_4 e^{S_2t}(1+\frac{V+p}{2\epsilon }),\nonumber\\
K_S(t)=C_3e^{S_1t}(1-\frac{V-p}{2\epsilon }) +C_4 e^{S_2t}(1-\frac{V+p}{2\epsilon }),
\end{eqnarray}
where $C_3=C_1/2$, $C_4=C_2/2$.

Using initial conditions $K_L(0)=1$ and $K_S(0)=0$ we have
\begin{eqnarray}
C_3=\frac{1}{4p}(p+V-2\epsilon ),\nonumber\\
C_4=\frac{1}{4p}(p-V+2\epsilon ).
\end{eqnarray}
Finally
\begin{eqnarray}
K_S(t)=\frac{V}{2p}(e^{S_1t}-e^{S_2t}),\nonumber\\
K_L(t)= \frac{1}{2p}[(p-2\epsilon )e^{S_1t}+(p+2\epsilon )e^{S_2t}].
\end{eqnarray}

We have obtained the exact wave functions of $K_S$- and $K_L$-components. (The expressions for $K^0(t)$ and $\bar{K}^0(t)$ are given by (8).) In (11) one can put $V/p\approx 1-2\epsilon ^2/V^2\approx 1$ since $\epsilon $ is extremely small.

\section{Results}
The probability of finding $K^0_{S}$ or, what is the same, the probability of $K^0_{L}K^0_{S}$ transition is given by $\mid \!K_S(t)\!\mid ^2$. After cumbersome calculation we get
\begin{eqnarray}
\mid \!K_S(t)\!\mid ^2=R\left[e^{-{\rm Im}(pt)}+e^{{\rm Im}(pt)}-e^{i{\rm Re}(pt)}-e^{-i{\rm Re}(pt)}\right]=
R\left[e^{-{\rm Im}(pt)}+e^{{\rm Im}(pt)}-2\cos({\rm Re}(pt))\right],\nonumber\\
R=\frac {1}{4}\mid V/p\mid ^2e^{{\rm Im}Vt+2{\rm Im}Mt}.
\end{eqnarray}
This expression is exact. $\mid \!K_S(t=0)\!\mid ^2=0$. From (11) it is seen that $\mid \!K_L(t=0)\!\mid ^2=1$.

Since $\epsilon $ is extremely small, in (12) we put $p\approx V+2 \epsilon ^2/V$ and assume $V/p=1$. Then
\begin{equation}
\mid \!K_S(t)\!\mid ^2=\frac{1}{4}e^{2{\rm Im}(Mt)}\left[e^{2\epsilon ^2\frac {{\rm Im}(Vt)}{\mid V\mid ^2}}+
e^{2{\rm Im}(Vt)-2\epsilon ^2\frac {{\rm Im}(Vt)}{\mid V\mid ^2}}-e^{-iVt-2i\epsilon ^2\frac {{\rm Re}(Vt)}{\mid V\mid ^2}}-e^{iV^*t+2i\epsilon ^2\frac {{\rm Re}(Vt)}{\mid V\mid ^2}}\right],
\end{equation}
To study the role of various terms, in (13) we put $\Gamma _{K^0}^d= \Gamma _{\bar{K}^0}^d=\Gamma ^d$, $m_{K^0}=m_{\bar{K}^0}=m$ ($m$ is the mass of $K^0$), ${\rm Re}U_{K^0}={\rm Re}U_{\bar{K}^0}$ and
\begin{eqnarray}
{\rm Im}U_{K^0}=-\frac{\Gamma _{K^0}^a}{2},\nonumber\\
{\rm Im}U_{\bar{K}^0}=-\frac{\Gamma _{\bar{K}^0}^a}{2},
\end{eqnarray}
where $\Gamma _{K^0}^a$ and $\Gamma _{\bar{K}^0}^a$ are the widths of absorption (not decay) of $K^0$ and ${\bar{K}^0}$, respectively. Then
\begin{eqnarray}
V=-i\frac{\Delta \Gamma }{2},\nonumber\\
\Delta \Gamma =\Gamma _{\bar{K}^0}^a-\Gamma _{K^0}^a. 
\end{eqnarray}
Equation (13) becomes
\begin{equation}
\mid \!K_S(t)\!\mid ^2=\frac{1}{4}\left[e^{-\frac {4\epsilon ^2t}{\Delta \Gamma }}+
e^{\frac {4\epsilon ^2t}{\Delta \Gamma }-\Delta \Gamma t}-2e^{-\frac{\Delta \Gamma t}{2}}\right]e^{-(\Gamma ^a_{K^0}+\Gamma ^d)t}.
\end{equation}

If $\Delta \Gamma t\gg 1$,
\begin{eqnarray}
\mid \!K_S(t)\!\mid ^2\approx \frac{1}{4}e^{-\Gamma (K_L\rightarrow K_S)t}e^{-(\Gamma ^a_{K^0}+\Gamma ^d)t},\nonumber\\
\Gamma (K_L\rightarrow K_S)=\frac{4\epsilon ^2}{\Delta \Gamma }=\frac{(\Delta m)^2}{\Delta \Gamma },
\end{eqnarray}
where $\Gamma (K_L\rightarrow K_S)$ is the width of $K^0_{L}K^0_{S}$ transition. The similar equation has been obtained in [14] for the width of $n\bar{n}$ transitions that is the verification of (17). (See also Refs. [10-13].)
 
Let us consider conditions where (16) is reduced to (17). For estimation, we assume that $\Delta \Gamma \approx \Gamma _{\bar{K}^0}^a$. Since $\Gamma t=L/l$, Eq. (17) holds when $\Gamma t\gg 1$ or
\begin{equation}
L\gg l,
\end{equation}
$l=1/N\sigma $. Here $L$ and $l$ are the any given distance and collision distance, respectively; $N$ is the number of nucleons in a unit of volume, $\sigma $ is the total cross section of $K^0N$ ($\bar{K}^0N$) interaction. For the copper plate and $\sigma =100$ mb we obtain $l\approx 2$ cm.

We would like to note the following. Since the Hamiltonian is non-hermitian, the field of applicability of potential description is restricted [11]. (For the oscillations in the external field [15-18] the Hamiltonian is hermitian and so there is no similar problem. We also note that Eq. (5) is consistent with Eqs. (1) and (2) in Ref. [15] in absence of external field.) For the problem under study the potential model can be used as a first approximation since the optical theorem (unitarity condition) is not used.

\section{Comparison with previous calculations}
The previous calculations (see Eqs. (7.83)-(7.89) of  Ref. [5]) give  \begin{eqnarray}
\left.\mid\!K'_L\right>_{st}=\left.\mid\!K_L\right>+r\left.\mid\!K_S\right>,\nonumber\\
r=i\pi N\Lambda f_{21}/[k(i\mu  +1/2)]\sim f_{21}/(i\Delta m/\Gamma _S+1/2),
\end{eqnarray}
$f_{21}=f-\bar{f}$, $\Delta m=m_L-m_S$, where $m_L$ and $m_S$ are the masses of stationary 
states, $f$ and $\bar{f}$ are the forward scattering amplitudes of $K^0$ and 
$\bar{K}^0$, respectively; $\Gamma _S$ is the decay width of $K_S^0$,
$\left.\mid \!K'_L\right>_{st}$ is the in-medium state expressed through the vacuum states. 
The notations are the same as in [5]. (See also Eqs. (1) and (2) of Ref. [3] or Eq. (9.32) of Ref. [4].)

For comparison with (17) one should construct the process width $\Gamma _{st}$ by means of (19). We have by definition
\begin{equation}
\Gamma _{st}=\mathcal{N}\int d\Phi \mid\!\left<\!K_S\mid\!K'_L\right>_{st}\mid ^2 \sim \mid \!r\!\mid ^2\sim \mid \! f_{21}/(i\Delta m/\Gamma _S+1/2)\!\mid ^2 \sim \mid \! f_{21}/\Delta m\!\mid ^2,
\end{equation}
where $\mathcal{N}$ is the normalization multiplier. Using the standard relation [5] between potential $V$ and forward scattering amplitude
\begin{equation}
V=U_{\bar{K}^0}-U_{K^0}=\frac{2\pi }{m}Nf_{21}
\end{equation}
and Eq. (15), we obtain
\begin{equation}
f_{21}=-\frac{im}{4\pi N}\Delta \Gamma .
\end{equation}
Finally
\begin{equation}
\Gamma _{st}\sim \left(\frac{m}{4\pi N}\right)^2\left(\frac{ \Delta \Gamma }{\Delta m}\right)^2\sim
\left(\frac{ \Delta \Gamma }{\Delta m}\right)^2. 
\end{equation}

Compared to (17), the $\Delta \Gamma $- and $\Delta m$- dependences are inverse. In other words, 
\begin{equation}
\Gamma _{st} \sim 1/\Gamma (K_L\rightarrow K_S). 
\end{equation}
We would like to stress that $\Delta \Gamma $ and $\Delta m$ are the crucial values which define the process speed.

We note that the comparison with [8] is non-trivial since conditions $\Gamma t\gg 1$ and the main requirement of perturbation theory are opposite. One should start from general Eq. (12). We will revert to this problem in the next paper. 

\section{Conclusion}
In the old model [1-5] the off-diagonal mass term was omitted. As shown above, this is a fundamental defect since it leads to a qualitative disagreement in the results. This means that regeneration has been not described at all. 

We have proposed an alternative model which is typical for the theory of the multistep processes [19]. The main part of this paper is exact solution of equations of motion. The result (12) is exact. Equations (16) and (17) depend on $\Delta m$ and the parameters of optical potentials which are defined from other problems. Their further uses have no need of a commentary as opposed to parameterization in the old calculation. We will continue our consideration in the following paper.


\newpage
\begin{thebibliography}{99}
\bibitem{1}
K.M. Case, Phys. Rev. {\bf 103}, 1449 (1956).
%
\bibitem{2}
M.L. Good, Phys. Rev. {\bf 106}, 591 (1957).
%
\bibitem{3}
M.L. Good, Phys. Rev. {\bf 110}, 550 (1958).
%
\bibitem{4}
T.D. Lee and C.S. Wu, Annu. Rev. Nucl. Sci. {\bf 16}, 511 (1966).
%
\bibitem{5}
E.D. Commins and P. H. Bucksbaum, {\em Weak Interactions of Leptons and Quarks}
(Cambridge University Press, 1983).
%
\bibitem{6}
F. Benatti, R. Floreanini, R. Romano, Phys. Rev. D {\bf 68}, 094007 (2003).
%
\bibitem{7}
G. Amelino-Camelia et al., Eur. Phys. J. C {\bf 68}, 619 (2010).
%
\bibitem{8}
V.I. Nazaruk, arXiv: 1510.01629 [hep-ph].
%
\bibitem{9}
V.I. Nazaruk, Phys. Lett. B {\bf 337}, 328 (1994).
%
\bibitem{10}
V.I. Nazaruk, Eur. Phys. J. C {\bf 53}, 573 (2008).
%
\bibitem{11}
V.I. Nazaruk, Int. J. of Mod. Phys. E {\bf 21}, 1250056 (2012).
%
\bibitem{12}
P.G.H. Sandars, J. Phys. {\bf G6}, L161 (1980).
%
\bibitem{13}
K.G. Chetyrkin et al., Phys. Lett. B {\bf 99}, 358 (1981). 
%
\bibitem{14}
B.O. Kerbikov, M.S. Lukashov, Y.A. Kamyshkov, L.J. Varriano, arXiv:1512.03398 [hep-ph]. 
%
\bibitem{15}
Xian-Wei Kang, Hai-Bo Li and Gong-Ru Lu, Phys. Rev. D {\bf 81}, 051901 (R) (2010). 
%
\bibitem{16}
H. B. Li and M. Z. Yang, Phys. Rev. D {\bf 74}, 094016 (2006).
%
\bibitem{17}
X. W. Kang, H. B. Li, G. R. Lu, arXiv: 1008.2845 [hep-ph].
%
\bibitem{18}
Yong-Feng Liu , Xian-Wei Kang , J. Phys. Conf. Ser. 738 (2016) no.1, 012043.
%
\bibitem{19}
V.I. Nazaruk, Eur. Phys. J. A {\bf 39}, 249 (2009).  

\end{thebibliography}
\end{document}